\documentclass[journal]{IEEEtran}

\usepackage{graphicx,cite,epsfig}
\usepackage{amsmath} 
\usepackage{amssymb}  
\DeclareMathOperator *{\argmin}{argmin}
\usepackage{pifont}
\usepackage{stmaryrd}
\usepackage{bbding}
\usepackage{subfigure}
\usepackage{algorithm}
\usepackage{algorithmic}
\usepackage{array}
\usepackage{amsthm}
\usepackage{multirow}
\usepackage{url}
\usepackage{mathrsfs}
\usepackage{color, soul}
\usepackage{mathrsfs}
\usepackage{float}

\newcommand{\tabincell}[2]{\begin{tabular}{@{}#1@{}}#2\end{tabular}}

\usepackage{booktabs}
\usepackage{threeparttable}
\usepackage{amssymb}

\begin{document}
\title{Toward Terabits-per-second Communications: Low-Complexity Parallel Decoding of $G_N$-Coset Codes}
\author{\IEEEauthorblockN{Xianbin~Wang, Jiajie~Tong, Huazi~Zhang, Shengchen~Dai, Rong~Li,  Jun~Wang} \\
	\IEEEauthorblockA{\IEEEauthorblockA{Hangzhou Research Center, Huawei Technologies, Hangzhou, China}\\
		Emails: \{wangxianbin1,zhanghuazi,lirongone.li,justin.wangjun\}@huawei.com}}
	
\maketitle
\thispagestyle{empty}

\begin{abstract}
Recently, a \emph{parallel} decoding framework of $G_N$-coset codes was proposed.
High throughput is achieved by decoding the independent component polar codes in parallel.
Various algorithms can be employed to decode these component codes, enabling a flexible throughput-performance tradeoff.
In this work, we adopt SC as the component decoders to achieve the highest-throughput end of the tradeoff.
The benefits over soft-output component decoders are reduced complexity and simpler (binary) interconnections among component decoders.
To reduce performance degradation, we integrate an error detector and a log-likelihood ratio (LLR) generator into each component decoder.
The LLR generator, specifically the damping factors therein, is designed by a genetic algorithm.
This low-complexity design can achieve an area efficiency of $533Gbps/mm^2$ under 7nm technology.
\end{abstract}

\section{Introduction}\label{section_introductions}
\subsection{$G_N$-coset codes}
$G_N$-coset codes, defined by Ar{\i}kan in~\cite{ArikanPolar}, are a class of linear block codes with the generator matrix $G_N$.

$G_N$ is an $N \times N$ binary matrix defined as
\begin{equation}\small\label{me_all}
\begin{aligned}
G_N \triangleq F^{\otimes n}
\end{aligned},
\end{equation}
in which $N=2^n$ and $F^{\otimes n}$ denotes the $n$-th Kronecker power of $F=[\begin{smallmatrix}
1 & 0  \\
1 & 1
\end{smallmatrix} ]$.

The encoding process is
\begin{equation}\small\label{me_all1}
\begin{aligned}
x_1^N = u_1^NG_N,
\end{aligned}
\end{equation}
where $x_1^N \triangleq \{x_1,x_2,\cdots,x_N\}$ and $u_1^N \triangleq \{u_1,u_2,\cdots,u_N \}$ denote the code bit sequence and the information bit sequence respectively.

An $(N,K)$ $G_N$-coset code~\cite{ArikanPolar} is defined by an information set $\mathcal{A}\subset\{1,2,...,N\}$, $|{\cal A}| = K$.
Its generator matrix $G_N(\mathcal{A})$ is composed of the rows indexed by $\mathcal{A}$ in $G_N$.
Thus (\ref{me_all1}) is rewritten as
\begin{equation}\small\label{me_all}
\begin{aligned}
x_1^N = u(\mathcal{A})G_N(\mathcal{A}),
\end{aligned}
\end{equation}
where $u(\mathcal{A})\triangleq\{u_i | i\in \mathcal{A}\}$.

The key to constructing $G_N$-coset codes is to properly determine an information set $\mathcal{A}$.
RM codes\cite{RM} and polar codes\cite{ArikanPolar} are two well-known examples of $G_N$-coset codes.
They determine $\mathcal{A}$ according to Hamming weight and sub-channel reliability, respectively.
\begin{figure}[]
	\centering
	\includegraphics[width=3.5in]{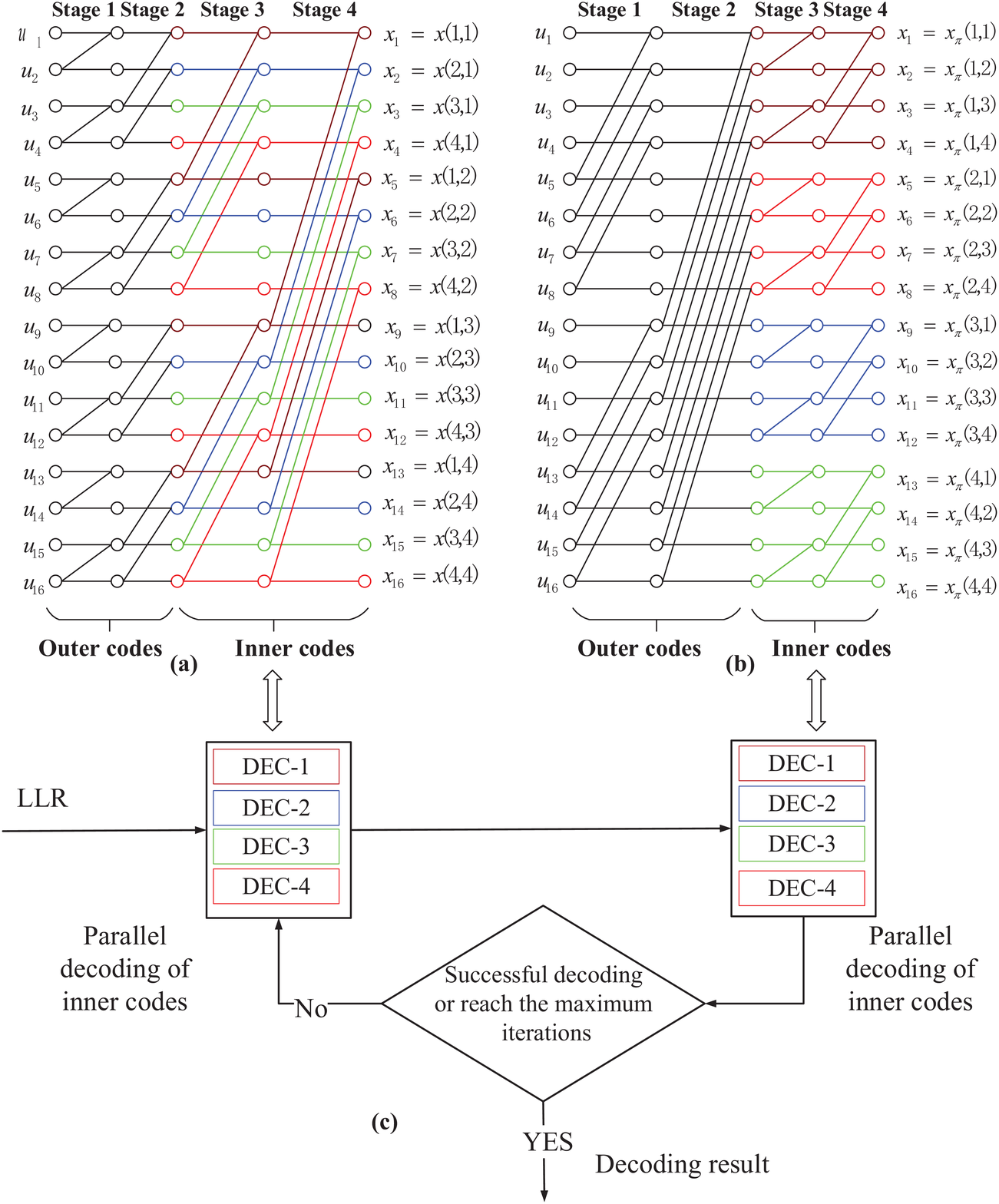}
	\caption{For $G_N$-coset codes, equivalent encoding graphs may be obtained based on stage permutations: (a) Ar{\i}kan's original encoding graph~\cite{ArikanPolar} and (b) stage-permuted encoding graph. Each node adds (mod-2) the signals on all incoming edges from the left and sends the result out on all edges to the right. (c) The parallel decoding framework only processes the inner code parts of the equivalent graphs, leaving the outer codes unprocessed. }
	\label{permutation}
\end{figure}

Recently, a parallel decoding framework of $G_N$-coset codes is proposed in~\cite{TPC}.
As shown in Fig.~\ref{permutation}(a), the encoding process of $G_N$-coset codes can be described by an $n$-stage encoding graph.
The former and latter stages respectively correspond to outer and inner codes.
The inner codes are \emph{independent} component codes that can be decoded in parallel~\cite{Syndrome}.

This parallel framework first produces \emph{equivalent encoding/decoding graphs} by permuting the inner and outer parts of the original decoding graph $\cal G$ (see Fig.~\ref{permutation}).
During decoding, we only process the inner code parts of these equivalent graphs, leaving the outer codes unprocessed.
The LLRs from different graphs about the same code bit are exchanged iteratively to reach a consensus.
Since all inner codes are decoded in parallel, this decoding framework supports a very high degree of parallelism.
The code construction under the parallel decoding algorithm is different from polar/RM codes, and is studied separately in~\cite{TPC}.

\subsection{Motivations and Contributions}
This paper mainly focuses on further enhancing decoding throughput.
The aforementioned iterative LLR exchange procedure requires soft-output component decoders such as SCL and SCAN to provide extrinsic LLRs~\cite{TPC}.
But if we aim at an ultra-high-throughput decoder, implementing soft-output component decoders gives rise to two problems.
First, the area efficiency of SCL and SCAN is much lower than fast-SC.
Second, the interconnections among the large number of component decoders consume considerable chip area.

To alleviate both problems, we propose to adopt hard-output SC as the component decoder.
First, the complexity and storage are reduced within each component decoder.
Compared with SCAN with one iteration, SC has $1/4$ decoding complexity and $1/2$ storage.
Compared with soft-output SCL with list size 8, SC has $1/16$ decoding complexity and $1/8$ storage~\cite{FinFET}.
Second, the interconnections among component decoders are also reduced.
Compared with soft-output decoders, hard-output SC decoders significantly simplifies routing because only hard bits are propagated among component decoders.

Besides, we introduce an error detector before each SC decoder to opportunistically reduce computation. If no
error is detected, we skip the SC decoding and directly output the hard decisions.

To minimize the performance loss due to the above simplifications, we propose a genetic algorithm based LLR generator.
LLR input (for this iteration) is generated from SC decoding output (from previous iterations) via a set of damping factors to determine the amplitudes.
The damping factors have a significant impact on the decoding performance, and is ``learned'' offline through a genetic algorithm based on unsupervised learning.
Compared with ``hand-picked'' parameters based on greedy stepwise optimization, the proposed genetic algorithm exhibits better performance.

\section{Stage permuted parallel decoding}
$G_N$-coset codes \cite{TPC} natively support parallel decoding, as the inner codes are independent.
To decode these component codes, various soft-output decoders, e.g., SCL, SC permutation list and SCAN, are employed \cite{TPC}.
In this work, we propose hard-output SC decoders to achieve higher area efficiency.
\begin{algorithm}[htb]
	\caption{Parallel decoding framework.}
	\label{alg:stage-permute-decoder}
	\begin{algorithmic}[1]
		\REQUIRE ~~\\
		The received signal $\mathbf{y} = \{y_i, i=1\cdots N\}$;\\
		\ENSURE ~~\\ 
		The recovered codeword: $\hat{\mathbf{x}} = \{\hat{x_i}, i = 1 \cdots N\}$;\\
		\STATE Initialize $L_{ch, i} \triangleq \frac{2y_i}{\sigma^2}, \forall i$;  $e_{\pi,i}=0, \ e_i=0, \ \forall i$; $\Lambda = \cal G$; \\
		\FOR {iterations: $t=1 \cdots t_{\max}$}
		\STATE Select decoding graph: $\Lambda = ~ {\cal G}_\pi~\text{if}~\Lambda ==\cal G~\text{else}~{\cal G}$;
		\IF {$\Lambda$ is ${\cal G}$}
		\FOR{inner component codes: $i=1 \cdots \sqrt{N}$ (in parallel)}		
		\STATE  $e_i = ErrorDetector( \hat{x}_{\pi,i,\forall j}^{t-1})$;
		\IF {$e_i==0$}
		\STATE $\hat{x}_{i,\forall j}^{t} = \hat{x}_{\pi,i,\forall j}^{t-1}$;
		\ELSE
		\STATE  $L_{i,j}^t = LLRgen (L_{ch, i + (j-1)\sqrt{N}}, \hat{x}_{\pi,i,j}^{t-1}, \hat{x}_{i,j}^{t-2}, e_{\pi,j})$, $\forall j$;	
		\STATE  $\hat{x}_{i,\forall j}^{t}=SCdecoder(L_{i,\forall j}^t)$;
		\ENDIF
		\ENDFOR
		\ELSE
		\FOR{inner component codes: $i=1 \cdots \sqrt{N}$ (in parallel)}
		\STATE $e_{\pi, i} = ErrorDetector(\hat{x}_{\forall j,i}^{t-1})$;	
		\IF {$e_{\pi, i}==0$}
		\STATE $\hat{x}_{\pi,\forall j, i}^{t} = \hat{x}_{\forall j,i}^{t-1}$;
		\ELSE
		\STATE $L_{\pi, j, i}^t = LLRgen (L_{ch, (i-1)\sqrt{N}+j}, \hat{x}_{j,i}^{t-1}, \hat{x}_{\pi,j,i}^{t-2}, e_j)$, $\forall j$;	
		\STATE $\hat{x}_{\pi,\forall j, i}^{t} = SCdecoder(L_{\pi,\forall j,i}^t)$;
		\ENDIF
		\ENDFOR
		\ENDIF
		\ENDFOR
	\end{algorithmic}
\end{algorithm}

\subsection{Parallel decoding framework}
\label{section_decoding}
The parallel decoding framework in \cite{TPC} is modified to support SC component decoders. In Algorithm~\ref{alg:stage-permute-decoder}, a $G_N$-coset code is alternately decoded on two factor graphs $\cal{G}$ and $\cal{G}_{\pi}$, as shown in Fig.~\ref{permutation}. The stage permuted graph $\cal{G}_{\pi}$ is generated by swapping the inner codes and outer codes in $\cal{G}$. Only the inner codes of each graph $\Lambda \in \{\cal{G},\cal{G}_{\pi}\}$ are decoded. And their decoding outputs are exchanged between the decoding graphs. The $\sqrt{N}$ component decoders can be implemented in parallel.

Now that we use SC to decode the component codes, the hard output must be converted into soft LLR as input for the next iteration.
Therefore, an LLR generator is placed before the SC decoder (line~8).
Meanwhile, an error detector is placed before the LLR generator (line~6).

For decoding graph $\cal G$ (resp. ${\cal G}_\pi$), the $j$-th code bit of the $i$-th inner component code is denoted by $x(i,j)$ (resp. $x_\pi(j,i)$).
Take graph $\cal G$ for example, the hard outputs (HO) from different component decoders of the previous iteration are combined into $\hat{x}_{\pi,i,\forall j}^{t-1}$ and then sent for error detection (line~6).
\begin{itemize}
  \item If no error is detected, i.e., the error detection output (E) $e_i=0$, then $\hat{x}_{\pi,i,\forall j}^{t-1}$ are directly taken as the new ``HO'' result of this iteration (line~8), and SC decoding is skipped.
  \item Otherwise, if $e_i=1$, the LLR of code bit $x(i,j)$ in the $t$-th iteration, denoted by $L_{i,j}^t$, is generated from channel LLR $L_{ch,i + (j-1)\sqrt{N}}$ and previous ``HO\&E'' results (line~10). The generated LLRs are decoded by SC to output new ``HO'' results (line~11).
\end{itemize}
Either way, new ``HO\&E'' results are sent to the next iteration.

After $t_{\max}$ iterations, the algorithm outputs the estimated codeword of the last decoding iteration as results.

\subsection{SC as component decoder}
\begin{figure}[]
	\centering
	\includegraphics[width=3.5in]{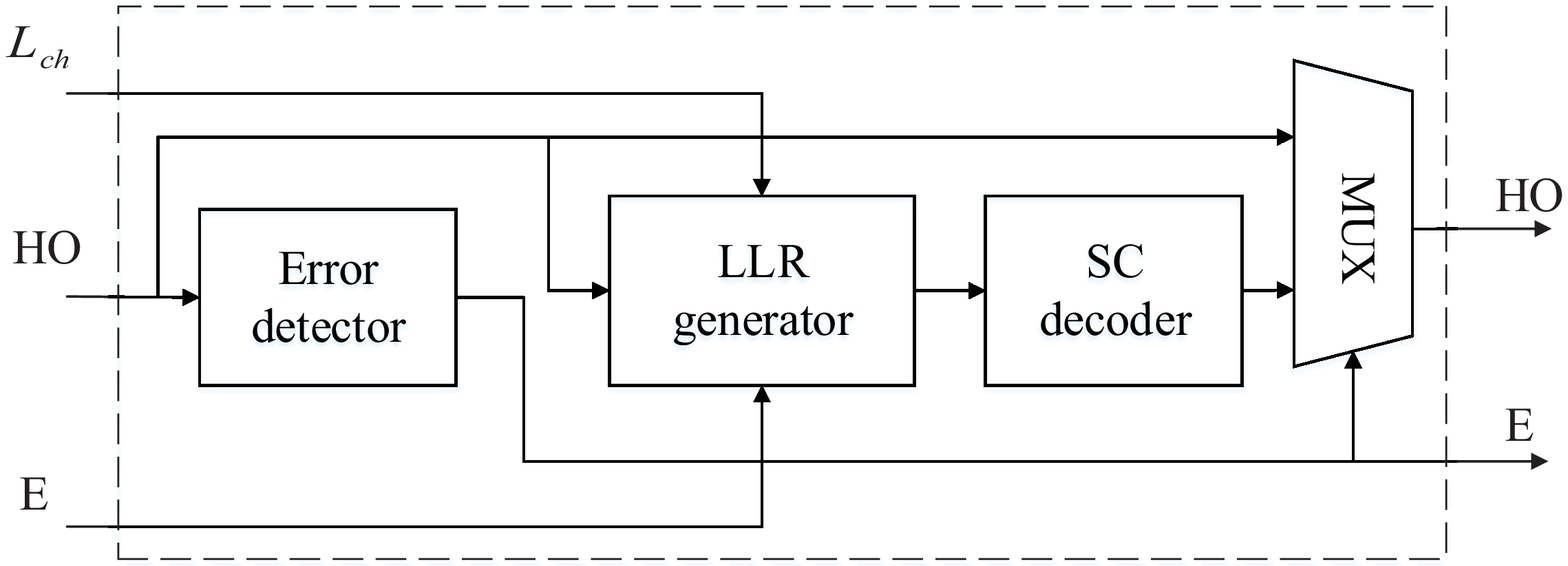}
	\caption{A component decoder consists of an error detector, an LLR generator and an SC decoder. The ``E'' result of this iteration is taken to switch the MUX.
		If no error is detected (E=0), the ``HO'' result from the previous iteration is directly taken as the new ``HO'' result of this iteration. Otherwise, LLRs are generated for SC decoding to output new ``HO'' result.}
	\label{whard}
\end{figure}
A component decoder consists of three parts, an error detector, an LLR generator and an SC decoder (see Fig.~\ref{whard}). An SC decoder takes soft LLR input but generates hard bits output.
The mismatch between hard output and soft input poses a challenge for iterative decoding, as the hard output cannot be directly used as soft input for the next iteration.
To solve this problem, an LLR generator is required to generate soft values from the hard output.

An error detector is placed before the LLR generator and it serves two purposes. First, if its input vector (i.e., HO from the previous iteration) is a codeword (no error detected), LLR generation and SC decoding can be skipped to save computation. Second, it provides a way to estimate the reliability of hard bits. Heuristically, if an input vector is already a codeword, they are deemed more reliable. Otherwise, if error detection failed, there is a chance that the error can not be corrected by an SC decoder, which implies less reliability. Therefore, error detection results facilitate the ``recovery'' of soft LLRs for the next iteration.

In practice, an error detector based on syndrome check can be implemented by reusing the encoding circuit. It costs almost no additional hardware resource.

The LLR generator is activated when an error detection fails.
It takes four inputs (i) the channel LLR, (ii,iii) the hard outputs of the previous two iterations, and (iv) the error detection output of the previous iteration.

Take the non-permuted graph $\cal G$ for example.
For code bit $x_{i,j}$, its input LLR is generated based on the previous-iteration error detection output $e_{\pi,j}\in\{0,1\}$.

If $e_{\pi,j}=1$, meaning error detection failed and hard output is from an SC decoder, the input LLR is the sum of channel LLR and hard outputs from the previous two iterations:
\begin{equation}\label{LLR_fail}
L_{i,j}^t = L_{ch, i + (j-1)\sqrt{N}} + \frac{2{\alpha}_t}{\sigma^2}  (1-2\hat{x}_{\pi,i,j}^{t-1}) - \frac{2{\beta}_t}{\sigma^2} (1-2\hat{x}_{i,j}^{t-2}),
\end{equation}
where $\alpha_t$ and $\beta_t$ respectively denote the damping factors, which determine the amplitude.

If $e_{\pi,j}=0$, meaning hard output is directly from an error detector since no error was found, the input LLR is the sum of the channel LLR and hard output from the previous iteration:
\begin{equation}\label{LLR_pass}
L_{i,j}^t = L_{chan, i + (j-1)\sqrt{N}} + \frac{2{\gamma}_t}{\sigma^2} (1-2\hat{x}_{\pi,i,j}^{t-1}),
\end{equation}
where the damping factor is denoted by $\gamma_t$.

Finally, the input LLR vector is sent to an SC decoder to output new ``HO'' results.

\section{Genetic algorithm based LLR generator design}\label{section_dampingfactors}
The LLR generator, parameterized by the three damping factors, has a significant impact on the overall performance.
Unfortunately, a theoretical optimum is difficult to obtain due to the following reasons.
First, the extrinsic information transfer analysis is hard with the proposed component decoder.
Second, the output of the component decoder is correlated with all its input vector due to the loopy decoding graph.
Both make conventional density evolution methods inapplicable.

Artificial intelligence provides an alternative method in the case where a precise theoretical approach is unavailable. Recently, deep learning, reinforcement learning and genetic algorithm have been applied to design better code constructions~\cite{AICoding} and decoding algorithms~\cite{AIflip,RLcoding}.

Inspired by this, we exploit a genetic algorithm based on unsupervised learning to design the damping factors. Damping factors play a similar role of chromosomes in the genetic algorithm, because they both individually and collaboratively contribute to the fitness of a candidate. A good candidate requires that all its damping factors are respectively good.
As such, a pair of good parents is likely to produce a good offspring, and this suggests that the genetic algorithm may ultimately converge to a good candidate.

At first, we start the genetic algorithm by initializing a population of size $M$. Each candidate contains $3t_{max}$ damping factors, including $\alpha_t$,  $\beta_t$  and $\gamma_t$, $t = 1,2,...,t_{max}$. $t_{max}$ denotes the maximum decoding iteration. We initialize each candidate as follows.
\begin{itemize}
	\item  Without any given prior knowledge, the initial damping factors are sampled from a uniform distribution $\mathcal{U}(0,v_\text{sup})$.
		By adjusting the parameter $v_\text{sup}$, we can trade optimality (a larger $v_\text{sup}$) for convergence rate (a smaller $v_\text{sup}$).
\end{itemize}
We observe that $\alpha_1$, $\beta_1$ and $\gamma_1$ (used to calculate LLR for the first decoding iteration) can be directly set to $0$ without any performance loss, since there is no information from the previous iteration. Similarly, $\beta_2$ is directly set to $0$.

The population are evaluated through Monte Carlo method and then ordered based on decoding performance.
The minimum signal-to-noise ratio to achieve a target block error rate (SNR@targetBLER) is taken as the performance metric.

Then, the algorithm enters a loop consisting of four steps.
\begin{enumerate}
	\item \textbf{Select} two distinct parents from the population. The $i$-th candidate is selected according to a probability $\frac{e^{-\lambda i}}{\sum_{j=1}^Me^{-\lambda j}}$ (normalized), where $\lambda$ is called the sample focus. In this way, a better candidate will be selected with a higher probability. By adjusting the parameter $\lambda$, we can
	tradeoff between exploitation (a larger $\lambda$) and exploration (a smaller $\lambda$).
	\item \textbf{Crossover} between parents to produce an offspring. Specifically, each damping factor of the offspring is randomly selected from the corresponding ones of its parents.
	\item \textbf{Mutate} the offspring randomly. This is implemented by independently mutating each damping factor with probability $p_\text{mutate}$.
	Specifically, if one damping factor is mutated, a random value sampled from Gaussian distribution $\mathcal{N}(0,\sigma_\text{mutate}^2)$ is added up to it.
	By adjusting $p_\text{mutate}$ and $\sigma_\text{mutate}^2$,  we can tradeoff optimality (larger $\sigma_\text{mutate}^2$ and $p_\text{mutate}$) and convergence rate (smaller $\sigma_\text{mutate}^2$ and $p_\text{mutate}$).
	\item \textbf{Insert} the offspring back to the population according to the decoding performance.
\end{enumerate}

The algorithm loop is terminated after reaching a maximum number of iterations.

\section{Performance evaluation}
\begin{figure}[]
	\centering
	\includegraphics[width=3.5in]{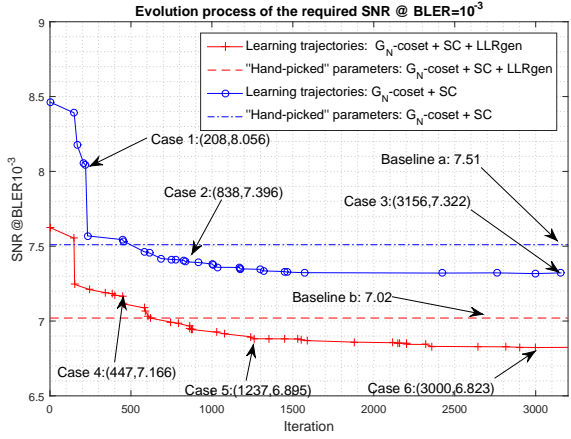}
	\caption{The learning trajectories of the SNR@BLER$=10^{-3}$.
	After about 600 iterations, the genetic algorithm ``learned'' better damping factors than the ``hand-picked'' ones based on greedy stepwise optimization.
	After about 3000 iterations, the algorithms in both cases converge.} 
	\label{fig_learningprocess}
\end{figure}

We evaluate the performance gain brought by the proposed LLR generator and the genetic algorithm, respectively.
The hyper parameters\footnote{The further optimization of the hyper parameters may bring improved performance. This is outside the scope of this paper.} of the genetic algorithm are provided in Table~\ref{hyper}.
\begin{table}
	\renewcommand{\arraystretch}{1.15}
	\caption{Hyper Parameters of genetic algorithm}
	\label{hyper}
	\centering
	\begin{tabular}{|c|c|}
		\hline
		Parameters &  Value \\
		\hline
		Population Size ($M$) & 32 \\
		\hline
		$v_\text{sup}$ & 2\\
		\hline
		Sample focus($\lambda$)   & 0.01  \\
		\hline
		Mutate probability ($p_\text{mutate}$)  & 0.07   \\
		\hline
		Mutate variance ($\sigma_\text{mutate}$)  & 0.3   \\
		\hline
	\end{tabular}
\end{table}
The learning trajectories of the required SNR to achieve BLER=$10^{-3}$ are presented in Fig.~\ref{fig_learningprocess}.
It shows that the decoding performance first improves rapidly as the genetic algorithm iterates and then converges gradually.

\begin{table}
	\renewcommand{\arraystretch}{1.15}
	\caption{The damping factors designed by genetic algorithm}
	\label{dampingfactors}
	\centering
	\begin{tabular}{|c|c|c|c|}
		\hline
		& $a_i$ &  $\beta_i$ & $\gamma_i$  \\
		\hline
		$i=1$ & 0 & 0 & 0  \\
		\hline
		$i=2$ & 0.2680 & 0 & 1.9997  \\
		\hline
		$i=3$ &  0.4236 & 0.2075 & 0.6695  \\
		\hline
		$i=4$ & 0.5051 & 0.2542 & 0.8296  \\
		\hline
		$i=5$ & 0.6147 & 0.3574 & 0.7598  \\
		\hline		
		$i=6$ &  1.2661 & 0.9922 &  0.7647  \\
		\hline		
		$i=7$ &  0.4054 & 0.2714 & 0.7851  \\
		\hline
		$i=8$ &  0.5360 &  0.1566 & 0.8723  \\
		\hline
	\end{tabular}
\end{table}

Two types of gains can be observed from Fig.~\ref{fig_learningprocess}.
First, the gain brought by the proposed LLR generator is $0.5$dB at BLER$=10^{-3}$, for both converged and non-converged cases.
The error detection results facilitate the ``recovery'' of soft LLRs in the proposed LLR generator, leading to the observed gain.
 
\begin{figure}[]
	\centering
	\includegraphics[width=3.5in]{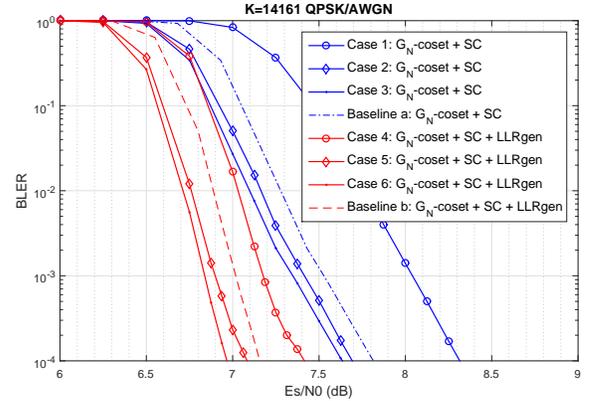}
	\caption{With converged damping factors, the gains brought by the proposed LLR generator and the genetic algorithm are $0.7$dB and $0.2$dB at BLER$=10^{-4}$.}
	\label{fig_learningprocessBLER}
\end{figure}

Second, we exemplify the ``learning gain'' through three points on the learning curve and present their BLER performances in Fig.~\ref{fig_learningprocessBLER}.
On the one hand, this proves the effectiveness of the genetic algorithm in designing good damping factors.
With the converged damping factors in Table~\ref{dampingfactors}, the proposed scheme is $0.2$dB better than the best ``hand-picked'' damping factors\footnote{The ``hand-picked'' method is a greedy stepwise optimization that chooses the best damping factors in every iteration.}.
On the other hand, it confirms that the component decoder with the proposed LLR generator exhibits better decoding performance than the case without it.

\begin{figure}[]
	\centering
	\includegraphics[width=3.5in]{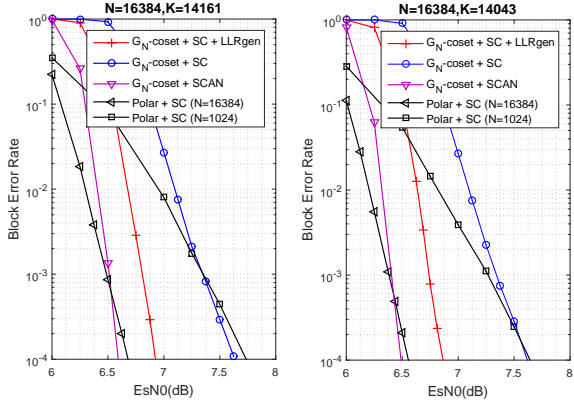}
	\caption{Compared with Type-1 and Type-2 baselines, the proposed decoder only trades $0.25$dB$\sim0.3$dB loss at BLER$=10^{-4}$ for improved area efficiency and reduced decoding latency.
		Compared with Type-3 baseline, the proposed scheme exhibits $0.75$dB gain at BLER$=10^{-4}$.
The polar codes are constructed by Gaussian approximation at Es$/$N0$=6.3$dB, $6.8$dB,  $6.0$dB and $6.8$dB for code rates $14161/16384$, $885/1024$, $14043/16384$ and $877/1024$, respectively.}
	\label{fig_float_simu}
\end{figure}

Next, we compare our scheme with some baselines in literatures.
\begin{enumerate}
	\item The same code construction decoded by the parallel soft output decoding algorithm~\cite{TPC}. This scheme exhibits a similar degree of parallelism to the proposed decoding algorithm, but incurs higher implementation complexity due to the difficulty in handling the internal decoder data flow.
	\item  A polar code with the same length and code rate, evaluated under SC decoding. It enjoys more coding gain but incurs larger decoding latency due to the serial nature of SC decoding.
	\item  A recently proposed polar coding scheme with similar target for terabit/s throughput~\cite{TbpsPolar}, which employs an unrolled hardware architecture for high throughput. ``Unrolling'' is only applicable for relatively short codes (e.g., $1024$) and thus sacrifices coding gain.
\end{enumerate}
The evaluation results are presented in Fig.~\ref{fig_float_simu}.
Compared with Type-1 and Type-2 baselines, the proposed decoder only trades $0.25$dB$\sim0.3$dB loss at BLER$=10^{-4}$ for improved area efficiency and reduced decoding latency.
Compared with Type-3 baseline, the proposed scheme exhibits $0.75$dB gain at BLER$=10^{-4}$.

\begin{figure}[]
	\centering
	\includegraphics[width=3.5in]{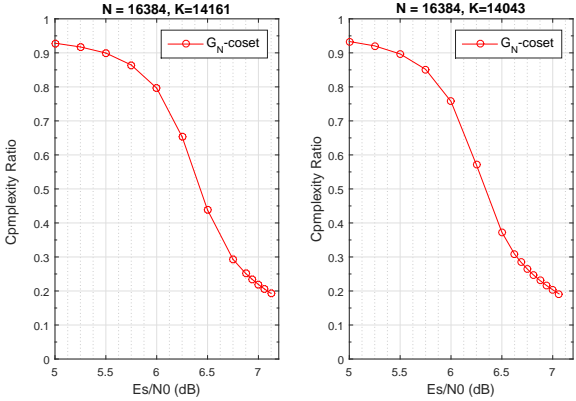}
	\caption{At BLER=$10^{-4}$, bypassing SC decoding can reduce $75\%$ decoding complexity.}
	\label{complexity}
\end{figure}

Then, we evaluate the complexity reduction due to skipped SC decoding.
The number of activated SC decoders is measured to evaluate the complexity. The results are presented in Fig.~\ref{complexity}.
It shows that the complexity reduction ratio varies with SNR.
For the case with higher SNR (lower BLER), more complexity is reduced.
At BLER=$10^{-4}$, bypassing SC decoding can reduce $75\%$ decoding complexity.

At last, the area efficiency of the proposed decoder is presented in Table \ref{table_throughput} (see details in our ASIC implementation \cite{Jiajie}).
With TSMC 16nm process, the area efficiency for code rate $14161/16384$ is $75Gbps/mm^2$ when the maximum number of iterations is eight.
The equivalent throughput under 7nm technology is about $322Gbps/mm^2$ with eight iterations and $533Gbps/mm^2$ with five iterations.
\begin{table}
	\renewcommand{\arraystretch}{1.15}
	\caption{Decoder Area Efficiency}
	\label{table_throughput}
	\centering
	\begin{tabular}{|c|c|c|c|c|c|}
		\hline
		Info & Iter- & Latency & Area Eff & \multicolumn{2}{c|}{Convert to}\\
		\cline{5-6}
		size &ation & (ns) &(Gbps/$mm^2$) &10nm &7nm \\
		\hline
		\multirow{4}*{\tabincell{c}{14161}}  
		& 5   &109.25	&120.73	&277.69	&533.16 \\
		\cline{2-6}
		& 6   &131.1	&100.61	&231.41	&444.30 \\
		\cline{2-6}
		& 7   &152.95	&86.24	&198.35	&380.83 \\
		\cline{2-6}
		& 8   &174.8	&75.46	&173.55	&322.22 \\
		\hline
	\end{tabular}
\end{table}

\section{Conclusions}
\label{section_conclusions}
In this work, we propose a low-complexity parallel decoding algorithm of $G_N$-coset codes.
The framework exploits two equivalent decoding graphs.
For each graph, the inner component codes are independent and support parallel decoding.
The component decoder adopts a novel design comprising an error detector, an LLR generator and an SC decoder.
The LLR generator, parameterized by a set of damping factors, is ``learned'' offline by a genetic algorithm based unsupervised learning.
The proposed decoding algorithm achieves comparable performance to the case with soft-output component decoder and conventional polar codes, but requires much lower decoding and hardware implementation complexity.

\bibliographystyle{IEEEtran}

\end{document}